\documentclass{article}

\usepackage{PRIMEarxiv}

\usepackage[utf8]{inputenc} 
\usepackage[T1]{fontenc}    
\usepackage{hyperref}       
\usepackage{url}            
\usepackage{booktabs}       
\usepackage{amsfonts}       
\usepackage{nicefrac}       
\usepackage{microtype}      
\usepackage{lipsum}
\usepackage{fancyhdr}       
\usepackage{graphicx}       
\graphicspath{{media/}}     
\usepackage{amsmath,amssymb,amsfonts}
\usepackage{algorithmic}
\usepackage{graphicx}
\usepackage{algorithm,algorithmic}
\usepackage{hyperref}
\hypersetup{hidelinks=true}
\usepackage{subcaption}
\usepackage{multirow} 
\title{Benchmarking ResNet for Short-Term Hypoglycemia Classification with DiaData
}

\author{
  Beyza Cinar\\
  Data Engineering\\
  Helmut Schmidt University\\
  Germany\\
  \texttt{cinarb@hsu-hh.de} \\
   \And
  Maria Maleshkova\\
  Data Engineering\\
  Helmut Schmidt University\\
  Germany\\
  \texttt{maleshkm@hsu-hh.de} \\\\
}

\begin{document}
\maketitle

\begin{abstract}
Individualized therapy is driven forward by medical data analysis, which provides insight into the patient's context. In particular, for Type 1 Diabetes (T1D), which is an autoimmune disease, relationships between demographics, sensor data, and context can be analyzed. However, outliers, noisy data, and small data volumes cannot provide a reliable analysis. Hence, the research domain requires large volumes of high-quality data. Moreover, missing values can lead to information loss. To address this limitation, this study improves the data quality of DiaData, an integration of 15 separate datasets containing glucose values from 2510 subjects with T1D. Notably, we make the following contributions: 1) Outliers are identified with the interquartile range (IQR) approach and treated by replacing them with missing values. 2) Small gaps ($\le$ 25 min) are imputed with linear interpolation and larger gaps ($\ge$ 30 and $<$ 120 min) with Stineman interpolation. Based on a visual comparison, Stineman interpolation provides more realistic glucose estimates than linear interpolation for larger gaps. 3) After data cleaning, the correlation between glucose and heart rate is analyzed, yielding a moderate relation between 15 and 60 minutes before hypoglycemia ($\le$ 70 mg/dL). 4) Finally, a benchmark for hypoglycemia classification is provided with a state-of-the-art ResNet model. The model is trained with the Maindatabase and Subdatabase II of DiaData to classify hypoglycemia onset up to 2 hours in advance. Training with more data improves performance by 7\% while using quality-refined data yields a 2-3\% gain compared to raw data.
\end{abstract}


\section{Introduction}
\label{sec:introduction}

Hypoglycemia is a life-threatening condition in patients with Type 1 Diabetes (T1D). Affected individuals can experience dizziness, loss of consciousness, and death due to a sudden decrease in glucose levels below 70 mg/dL. T1D is an autoimmune disease, characterized by insufficient insulin production, necessitating lifelong insulin replacement therapy \cite{idf2025}. Controlled glucose levels and consistent monitoring can decrease the risk of severe complications. Current treatment guidelines emphasize the integration of technological advances like continuous glucose monitoring (CGM) devices and insulin pumps, which can reduce the occurrence of severe hypoglycemia \cite{elsayed_14_2023}. CGM devices are minimally invasive sensors that continuously measure glucose levels in interstitial fluid. They enhance diabetes management and support personalized therapeutic adjustments \cite{jessica_lucier_type_2023}. When augmented with data analytics and artificial intelligence (AI), these devices can provide trend analysis or predict adverse events such as hypoglycemia \cite{Felizardo2021review}. Recent studies report that glucose prediction models can improve life quality and reduce fear of hypoglycemia, and diabetes-related distress~\cite{Ehrmann2024}.

Consequently, hypoglycemia prediction has emerged as a prominent area of research, aiming to support timely interventions. AI methods, particularly deep learning (DL) on time series data, show potential for early detection \cite{Felizardo2021review, sergazinov_glucobench_2024}. However, machine learning (ML) models require high-quality, large datasets to learn generalizable patterns and increase reliability. A major challenge poses the fact that the field currently lacks sufficiently large and heterogeneous CGM and sensor datasets specifically focused on individuals with Type 1 Diabetes. To address this gap, we previously proposed DiaData, a large integrated CGM dataset of 15 separate datasets, comprising 2510 subjects with T1D \cite{DiaData_Preprint}. 

This study advances our previous work and aims to improve the quality of DiaData, thus providing the foundation for data analysis, without misleading sensor errors. CGM devices and wearables measuring physiological data such as heart rate typically have the drawbacks of sensor errors, over- or underestimation, especially for hypoglycemic glucose ranges, and data gaps. Missing values are common and can arise due to sensor errors, human mistakes, during data transmission, lost connections, battery, or hardware-related issues \cite{Torkey2021, Rehman2024}. Therefore, data cleaning, anomaly detection, and missing value treatment are essential steps to ensure high-quality data, especially if diabetes management depends on ML-based models \cite{Rehman2024, Torkey2021, Cichosz2025}. 

We complement the quality-refined DiaData with a benchmark for hypoglycemia classification up to 2 hours before its onset. The benchmark validates the effectiveness of DiaData by ensuring that performance is reproducible, comparable, and potentially improved, while also testing generalization to unseen data. 
In particular, this study makes the following contributions: 1) DiaData enhanced with outlier pruning for glucose and heart rate data. 2) DiaData with imputed data gaps. Missing values are grouped into 5-25 minutes and 30-120 minutes to impute small gaps with linear and larger gaps with Stineman interpolation. Gaps exceeding 24 hours are not treated. 3) DiaData is preprocessed for the benchmark use case. Five classes are defined, corresponding to 0, 5–14, 15–29, 30–59, and 60–120 minutes before hypoglycemia. 4) Results of the correlation analysis between glucose and heart rate values for each of the five classes, focusing on the time intervals preceding hypoglycemia. 5) Finally, a benchmark on hypoglycemia classification up to 2 hours before onset is presented, training the quality-enhanced DiaData on a state-of-the-art ResNet model. 

The remainder of this study is structured as follows. Section \ref{sec:stoa} reviews methods for data cleaning and the performance of ML models for hypoglycemia classification. Section \ref{sec:methods} provides statistical insights into DiaData, presents the applied data cleaning and preprocessing methods, and reports the model architecture. Section \ref{sec:results} shows the results of the data analysis, section \ref{sec:discussion} discusses the results, and lastly, section \ref{sec:conclusion} concludes the study.

\section{Related Work}
\label{sec:stoa}
Continuous sensor data reveals insights into physiological relationships, enabling anomaly prevention and personalized treatment. However, they can introduce discrepancies and lead to significant information loss \cite{Cichosz2025}. 

For identifying outliers, various methods are proposed, such as the interquartile range (IQR) outlier detection, the 3 sigma method, linear models, multivariate clustering techniques, and the mean absolute deviation model \cite{Nurhaliza2024}. After reviewing, Nurhaliza et al. concluded that the IQR outlier detection method is superior and enhances the performance of the model, particularly for medical data. It could identify more outliers in total than the 3 sigma method \cite{Nurhaliza2024}. The IQR method estimates the unevenness while splitting data into quartile ranges. After identification, anomaly data can be removed, smoothed, or treated like missing values \cite{Torkey2021}. Torkey et al. applied the IQR method and replaced outliers with missing values, which were imputed with a Random Forest (RF) Regressor \cite{Torkey2021}.

Missing values are common and can be induced by various situations, including taking off the device for charging, replacing the device, or not correctly attaching the device \cite{Rashtian2021}. Moreover, sampling sensor data to the same frequency can increase missingness. Likewise, a variety of methods have been compared. Acuna et al. report that trained machine learning (ML) methods typically yield the best accuracy when using linear interpolation \cite{Acuna2023}. Cichosz et al. investigate different CGM imputation techniques for subjects with T1 and Type 2 diabetes. For gaps missing completely at random, it is reported that simple removal, RF, Piecewise Cubic Hermite Interpolating Polynomial (PCHIP), and linear interpolation achieve the best alignment. For most larger gaps, removal, and temporal alignment imputation (TAI) outperformed other methods \cite{Cichosz2025}. Jeon et al. highlight that the lowest root mean square error (RMSE) score is obtained with linear interpolation, Stineman interpolation, and Kalman smoothing with a structural model \cite{jeon_predicting_2020}. Furthermore, Rehman et al. investigate the performance of imputation methods while creating different gap durations to enable a comparison with true values. They compared linear, nearest neighbor, PCHIP, MAKIMA, splines, and previous interpolations. Linear and MAKIMA interpolations yielded better results, whereas MAKIMA was better for larger gaps \cite{Rehman2024}. 
Consequently, the performance of the better methods varies across studies and depends on the evaluation methods, utilized data, and prediction models. Usually, linear interpolation is superior, simple to implement, and has a short computation time, but is more often recommended for smaller gaps. For increased gaps, MAKIMA, Stineman, or ML-based interpolations could perform better. However, regression models like MICE and RF have a very high computation time.

Hypoglycemia is the leading cause of mortality in T1D. Thus, its prediction has become a prominent area of research, aiming to enable intervention following clinical guidelines, with AI methods showing potential for early detection. Particularly, time series-based deep learning (DL) methods are advantageous when being trained on large data \cite{Felizardo2021review, sergazinov_glucobench_2024}. 
In this context, Dave et al. classified hypoglycemia onset using a binary classification-based RF model for various PHs (15, 30, 45, and 60 min). They trained on data from 112 subjects and achieved high accuracy, with performance decreasing as the PH increased \cite{dave_feature-based_2021}. Results for precision (PRE), recall (RCL), and F1-Measure (F1-M) are presented in Table \ref{tab:CompStoA}.
D'Antoni et al. proposed a two-step classification framework using an LSTM model. The first model classified between euglycemia, hypoglycemia, and hyperglycemia. Then, the time to the event was classified, ranging from 15 to 120 min. The model, which was trained on the OhioT1DM dataset comprising 12 subjects, exhibited reduced performance due to class imbalance, particularly in terms of recall. Reliable predictions are primarily observed for 15- and 30-minute horizons \cite{dantoni_prediction_2022}.
Similarly, Cinar et al. developed a 1DCNN model to classify hypoglycemia up to 2 hours in advance. All PHs were integrated into a single multi-class classification model. The model, trained on glucose and heart rate data from 9 subjects and evaluated on a separate validation dataset, also showed declining performance with increasing PHs and overall instability, while recall remained comparable to precision \cite{xie_transfer_2024}. In a related study, Cinar et al. used the OhioT1DM dataset and incorporated multiple input features, including CGM, basal and bolus insulin, and acceleration magnitude, into an LSTM model. While the model aimed to predict hypoglycemia up to 4h in advance, strong performance was only achieved for the 0–15 minute window before onset \cite{cinarMaster}.
Finally, Hüni et al. employed both LSTM and Graph Attention Networks (GAT) for binary classification of hypoglycemia. For a 15-minute PH, LSTM performed better, while GAT yielded superior results at longer horizons. Thus, Table \ref{tab:CompStoA} presents the results of the GAT model. Both models yielded low precision and indicated high false alarm rates as well. Additionally, they report that longer ISL did not lead to better performance \cite{Hni2025}.
Class imbalance is a significant challenge, as the relevant classes represent rare conditions. Mahdy et al. suggest oversampling less present classes by duplicating the number of units of hypoglycemic values \cite{Mahdy2024}. Other methods include using synthetic data to simulate possible values, such as applying SMOTE or using ML approaches \cite{daniel_onwuchekwa_enhanced_2024}. Furthermore, data can be undersampled, which reduces overrepresented classes to the minority class.

\begin{table}[!ht]
\centering
\caption[Related Work for Hypoglycemia Classification]{Related Work for Hypoglycemia Classification \label{tab:CompStoA}}
\begin{tabular}{c |c |c |c|c |c |c |}
\cline{2-7}
    & \textbf{Metric}  &
    \multicolumn{5}{ l |} {\textbf{Class}}\\ 
    \cline{3-7} & & 0& 1  & 2 & 3 & 4  \\ \hline

\multicolumn{1}{|l|}{\multirow{4}{*}{}} & PRE  & N/A & 0.77 & 0.69 &0.45 &0.22  \\

\cline{2-7} 
\multicolumn{1}{|l|}{D'Antoni } & RCL   & N/A & 0.81 &0.55 &0.29&0.26    \\ 

\cline{2-7}
\multicolumn{1}{|l|}{et al., \cite{dantoni_prediction_2022}}  & F1-M   & N/A & 0.79& 0.61 &0.35 &0.22 \\
 \hline

\multicolumn{1}{|l|}{\multirow{4}{*}{}}&  SE  & N/A&0.98 &0.97& 0.96 & N/A  \\  

\cline{2-7} 
\multicolumn{1}{|l|}{Dave et al., \cite{dave_feature-based_2021}} & SPE & N/A & 0.98 & 0.95& 0.96 & N/A   \\ 
 \hline

\multicolumn{1}{|l|}{\multirow{4}{*}{}} &  PRE  &0.84 &0.43 &0.63& 0.59 & N/A  \\  

\cline{2-7} 
\multicolumn{1}{|l|}{Cinar et al., \cite{xie_transfer_2024}} & RCL   &0.80 &0.69 &0.57 &0.57 & N/A   \\ 

\cline{2-7}
\multicolumn{1}{|l|}{}  & F1-M  & 0.82 &0.53 &0.60& 0.58&N/A \\
 \hline

\multicolumn{1}{|l|}{\multirow{4}{*}{}} & PRE   & 0.95  & 0.47 &0.27&0.29 &0.37   \\  

\cline{2-7} 
\multicolumn{1}{|l|}{Cinar et al., \cite{cinarMaster}} & RCL   &  0.96 & 0.60 & 0.38&0.31& 0.23   \\ 

\cline{2-7}
\multicolumn{1}{|l|}{ }  & F1-M   &  0.95& 0.53  & 0.32 & 0.30 & 0.28  \\
\hline

\multicolumn{1}{|l|}{\multirow{4}{*}{}} & Precision & N/A & 0.13 &  0.15 & 0.14  & N/A  \\  

\cline{2-7}
\multicolumn{1}{|l|}{Hüni et al.,\cite{Hni2025}} & Recall   & N/A &  0.78 & 0.75&  0.77 & N/A    \\ 

\cline{2-7}
\multicolumn{1}{|l|}{}  & F1-M  & N/A &  0.39 & 0.41 &  0.40 & N/A  \\
 \hline

\end{tabular}
\end{table}

\section{Methodology}
\label{sec:methods}
This study improves the quality of the previously reported DiaData, a large integrated dataset of patients with T1D comprising CGM data \cite{DiaData_Preprint}. The aim is to enable a more reliable analysis of glucose variability within the dataset and across demographic groups. Furthermore, the improved dataset serves as a basis for a correlation analysis between sensor signals and for training a DL model to classify hypoglycemia onset. This section first provides insights into DiaData. Then, a data preprocessing framework is applied to increase data quality, addressing outliers and missing values, preprocessing the data for hypoglycemia classification, and applying data analysis methods on the quality-refined data.

\subsection{Dataset}

DiaData integrates 15 different datasets \cite{RodriguezLeon2023, DiatrendPaper, HupaMendeley, ShanghaiFigshare, ictinnovaties_diabetes_heart_rate_2025, jab_center_diabetes} into one large database containing CGM measurements in 5 min intervals. The database consists of a main dataset including the timestamps, patient IDs, and CGM values. Moreover, two subdatabases are extracted. Subdatabase I includes glucose and demographic data (age, age group, and sex). It comprises various age groups (0$-$2, 3$-$6, 7$-$10, 11$-$13, 14$-$17, 18$-$25, 26$-$35, 36$-$55, 56$+$). Age groups were created since the exact age was not available for each subject. Additionally, age groups enable cluster-based data analysis of newborns, children, teenagers, young adults, adults, and the elderly. Subdatabase II includes sensor data of CGM and heart rate. Further insights are provided in Table \ref{tab:datasetfusedsummary}. The hypoglycemic data points of the main database comprise only up to 4\% of all glucose values underlying the challenge in hypoglycemia prediction. Subdatabase II comprises only a small subset of the main database, which could lead to limitations for DL models, despite including heart rate data.
\begin{table}[ht]
\centering
\caption{Statistics of Integrated Datasets}
\label{tab:datasetfusedsummary}
\begin{tabular}{|p{2cm}|p{2cm}|p{2cm}|p{3cm}|p{3cm}|}
\hline
\textbf{Dataset} & \textbf{Subjects} & \textbf{Total Values} & \textbf{Values $\le70$ mg/dL} &  \textbf{Missing Values} \\
\hline
Main Dataset  & 2510 & 105 mil. & 4 mil. & 132 mil. \\
Sub-Dataset I  & 2475  & 105 mil. & 4 mil. &  131 mil. \\
Sub-Dataset II & 51   & 406372   & 18880 &  1 mil. \\
\hline
\end{tabular}
\end{table}

To integrate the individual datasets, semantic consistency, uniform sampling frequency, and alignment of sensor parameters were required. Accordingly, all glucose values were converted to mg/dL and into float values. Moreover, all datasets were resampled to 5-minute intervals, while five datasets were undersampled to 5-minute intervals to avoid reducing data volume by oversampling the remaining ten. In addition, for hypoglycemia prediction, differences between 5, 10, and 15 minutes can be clinically significant. Undersampling introduced small gaps that can be reliably imputed, however, oversampling would risk information loss, particularly for high-frequency signals like heart rate, which can change rapidly. In the previously published version of DiaData, missing values from undersampling were forward-filled to address artificial gaps without providing fully preprocessed data \cite{DiaData_Preprint}. In contrast, the present study did not impute artificial missing values during data integration. Instead, these values were deliberately left unprocessed until the data cleaning stage. This allowed for the application of more appropriate and robust data cleaning and imputation techniques, surpassing the previously used feedforward filling method. In particular, outliers were addressed with the IQR methods, small gaps were imputed with linear interpolation, and larger gaps with Stineman interpolation. 

\textbf{\textit{Outliers}}: Statistical analysis reveals discrepancies across the datasets that require data cleaning. It was previously identified that CGM devices often overestimate glucose levels, resulting in extreme outliers between 400–600 mg/dL \cite{DiaData_Preprint}. Conversely, hypoglycemic values are frequently underestimated, with some readings being below 30 mg/dL. These values may not be realistic and are known limitations of certain CGM devices. Moreover, anomalies distort the distribution and skewness of glucose values, introducing bias between single datasets. Additional errors may arise from environmental changes, false sensor placement, or device changes, further affecting data quality. Similar issues are observed in heart rate measurements with implausible values outside the range of 30-200 bpm. Notably, the D1NAMO dataset even reports heart rates exceeding 250 bpm, below 6 bpm, and contains a lot of zero values, which are not marked as outliers. These could indicate unworn sensor durations or failed measurements, likely replacing missing values with zeros. Such anomalies, which can be influenced by variability across sensor brands, affect data quality, leading to inconsistent distributions and risks of misleading model analysis and predictions.

\begin{figure}[htbp]
\centerline{\includegraphics[scale=0.4]{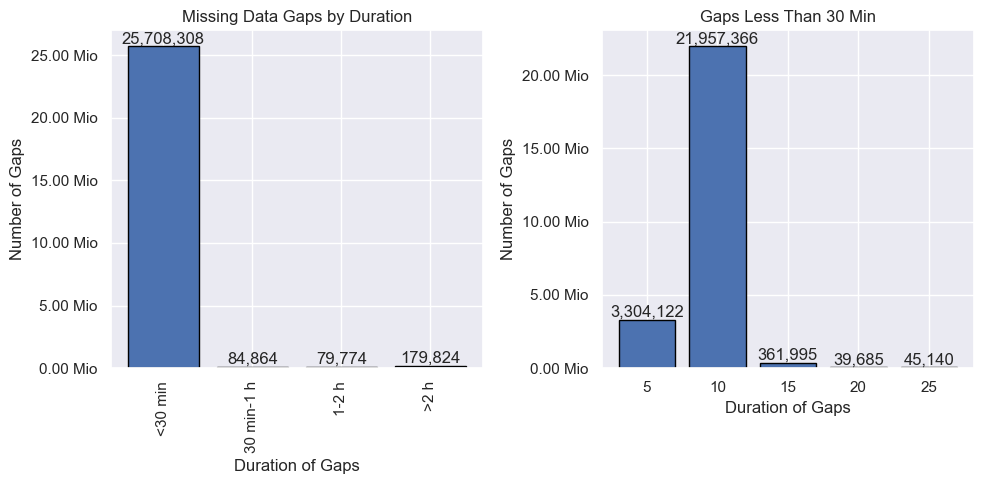}}
\caption{Missingness in Glucose Levels Before Data Imputation}
\label{fig:missingbeforeimp}
\end{figure}

\textbf{\textit{Missingness}}: DiaData contains a substantial amount of missing data, as can be seen in Table \ref{tab:datasetfusedsummary}. Furthermore, Fig. \ref{fig:missingbeforeimp} reveals that many of the gaps are short-term, under 30 min, partially introduced by undersampling. Although undersampled datasets exhibit increased missing data (T1DGranada, RT-CGM), these discrepancies may also result from a larger number of subjects or longer study durations. These short gaps can be imputed with simple but reliable methods, keeping a similar pattern of glucose values and reducing information loss.

\subsection{Quality Improvement} 
The data cleaning framework for DiaData includes the treatment of extreme and missing values to improve its quality and reduce bias and sensor error. First, outliers were determined with the IQR approach, identified as the superior method for glucose values in section \ref{sec:stoa}. Anomaly data was then replaced with missing values. Thereafter, all missing values were classified into gap sizes of less than 30 min, between 30 and 120 min, and more than 120 min, to impute different gap lengths with different techniques. 

\textit{\textbf{Outlier Removal:}} 
First, we replaced zero values in CGM and heart rate data with missing values, since these data points were included in the first (Q1) or second quartiles (Q2), which could limit outlier detection methods in recognizing low-value anomalies.
Then, we applied the IQR method adapted from Torkey et al. \cite{Torkey2021} separately for the Maindatabase, Subdatabase I, and Subdatabase II, to determine anomalies in glucose and heart rate data. For each subject, values falling outside of 1.5 times the interquartile range (IQR = Q3 – Q1) were masked as outliers, assuming a normal distribution. Additionally, given the known limitations of CGM accuracy in hypoglycemic ranges, values less than 40 mg/dL were masked \cite{Alva2023, Spartano2024}. We also removed values above 500 mg/dL \cite{Alva2023}. For heart rate, values above 200 bpm and below 30 bpm were removed. Finally, all masked outliers were replaced with missing values. Anomaly treatment standardized the variance of glucose measurements across the single dataset, enhancing its reliability for analysis. The most significant improvements were observed in the T1GDUJA dataset. Outlier treatment leads to a more realistic and harmonized distribution, especially in the D1NAMO dataset.

\textbf{\textit{Imputation:}} For imputing missing values, the gaps were classified by duration of short-term (less than 30 min), middle-term (between 30 and 120 min), and larger gaps (more than 2 hours). Section \ref{sec:stoa} has concluded that a single imputation method may not provide reliable results across all gap lengths, as also reported by Gupta et al. \cite{GuptaRev2025titlex}. Linear interpolation was identified as superior for short gaps. Thus, gaps shorter than 30 min were imputed linearly. Linear interpolation imputes a gap between two known values as a straight line \cite{noor_filling_2013}.

For data gaps longer than 30 minutes but not exceeding 2 hours, Stineman interpolation was applied, as also recommended by Jeon et al. \cite{jeon_predicting_2020}. This method fits a Stineman cubic curve between the endpoints of each gap. Stineman's algorithm employs a piecewise rational function to ensure smooth and stable interpolation across the missing interval \cite{jeon_predicting_2020}.

\begin{figure}[htbp]
\centerline{\includegraphics[scale=0.4]{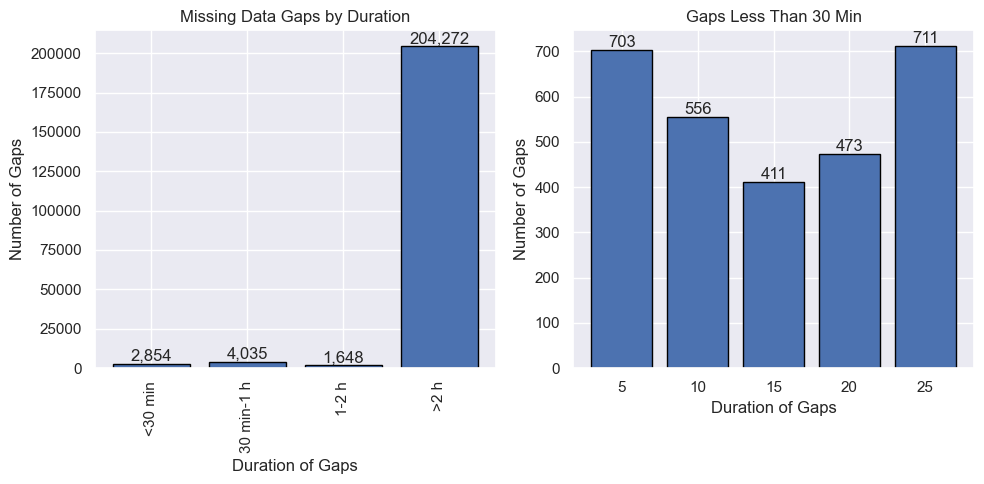}}
\caption{Missingness in Glucose Levels After Data Imputation}
\label{fig:missingafterimp}
\end{figure}

Lastly, Fig. \ref{fig:missingafterimp} summarizes that most missing values could be treated successfully, whereas some short-term and very few middle-term gaps remained, possibly because those could not provide sufficient known values. 
\begin{figure}[htbp]
\centerline{\includegraphics[scale=0.3]{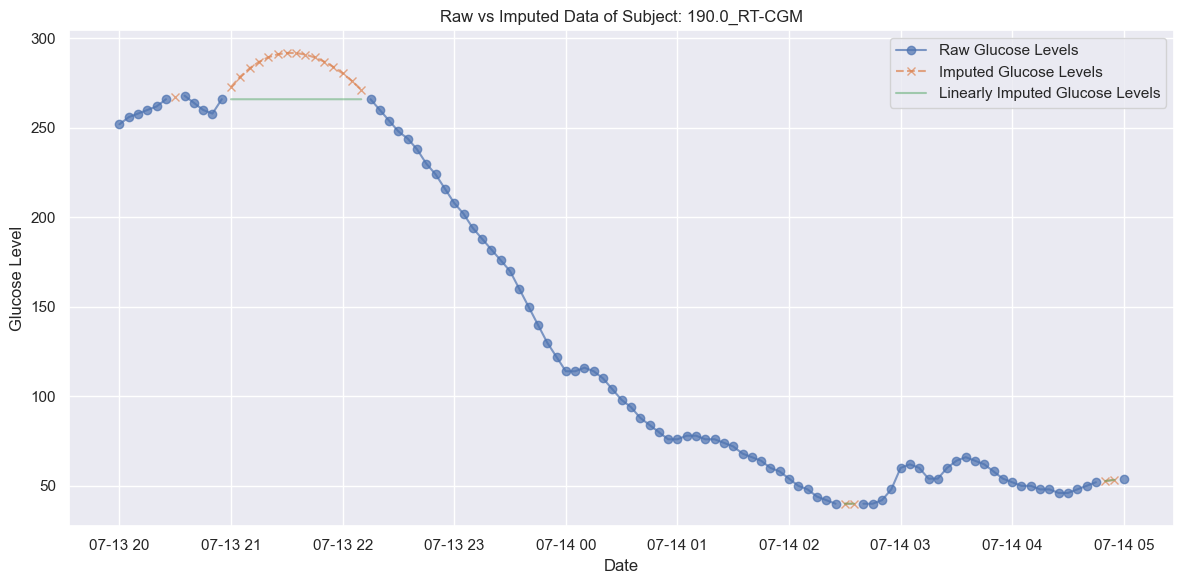}}
\caption{Raw vs. Imputed Data for Subject 190.0 RT-CGM.}
\label{fig:visualimputed1}
\end{figure}

In addition, all missing values were also imputed using linear interpolation to allow for visual comparison, as conventional validation metrics may fail to capture the true underlying patterns of the data, especially in the absence of ground truth values. Fig. \ref{fig:visualimputed1} presents 9 hours of data, in which the first missing values occur, for a randomly selected subject. It can be observed that linear interpolation fails to capture slopes and peaks, merely connecting known values linearly. This can lead to overestimation and information loss, misrepresenting true glucose dynamics over extended periods. However, Fig. \ref{fig:visualimputed1} illustrates reliable imputation for small gaps.

\subsection{Data Preprocessing}

Data preprocessing is a critical step in preparing datasets for analysis and ML, particularly for time series classification aimed at hypoglycemia prediction. First, class labels need to be assigned to the dataset, then sensor values should be normalized. Finally, time series can be created, and data imbalance needs to be addressed.

\textbf{\textit{Class Assignment:}} The benchmark use case focuses on hypoglycemia classification up to 2 hours before the event. Therefore, the hypoglycemic event was identified with a threshold ($\le$ 70) and assigned class 0. Other classes were computed backward from class 0. The distribution of classes representing time ranges before hypoglycemia is presented in Table \ref{tab:classesdistr} for the Maindatabase (MDB) and Subdatabase II (SDBII) and is mainly adapted from \cite{xie_transfer_2024, cinarMaster, daniel_onwuchekwa_time_2024}. 
\begin{table}[h!]
\centering
\caption{Distribution of Classes}
\label{tab:classesdistr}
\begin{tabular}{|c|c|c|c|c|c|}
\hline
\textbf{Class} & \textbf{0}& \textbf{1}&\textbf{2}  & \textbf{3} & \textbf{4}   \\
\hline
\textbf{Time range} & \textbf{0} & \textbf{0-15} & \textbf{15-30} & \textbf{30-60} & \textbf{60-120} \\
\hline
\textbf{MDB} & 6283444&1255676& 1775458& 3305367& 6485889 \\
\hline
\textbf{SDBII} & 39842 &  7416 &10726 &20174 &  39028  \\ \hline

\end{tabular}
\end{table}

\textit{\textbf{Normalization:}} After class assignment, values were normalized with min-max scaling, normalizing sensor data between zero and one. Normalization is essential for ML models sensitive to feature scale, as it leads to faster convergence during training \cite{Pujol2024}. Since a population-based model is aimed, data is normalized globally to mitigate overestimation and bias towards overrepresented subjects. Moreover, we aimed to keep the variance between subjects. Finally, the remaining missing values in the heart rate column were replaced with zero as suggested by Mahdy et al. \cite{Mahdy2024}.

\textbf{\textit{Time Series Generation:}} Thereafter, time series were generated using a sliding window approach, applied individually to each subject. Only the MDB and SDB II were used to enable performance comparison with prior work limited to only sensor values and to apply deep learning-based time series classification models, which do not incorporate demographic information. The ISL matched the maximum prediction horizon (PH) and was set to 2 hours, corresponding to 25 time points. Each window was required to span exactly 2 hours, contain no missing values or gaps, and was labeled according to the final instance in the sequence. Subjects lacking sufficient data for the defined window length were excluded, resulting in a final cohort of 2494 subjects in the MDB and 47 in SDB II. 
Data were then split into training, validation, and test sets using a 7:1.5:1.5 ratio \cite{Shao2024, Pujol2024}. To maintain temporal consistency and prevent future values from appearing in test data and past values in train data, splitting was performed chronologically for each subject without shuffling. First, the data were divided into temporary and test sets (8.5:1.5), and subsequently, the temporary set was split into training and validation sets using an 8.235:1.765 ratio.

\textbf{\textit{Class Imbalance:}} Table \ref{tab:classesdistr} presents the class distribution, revealing class imbalance in both databases. Especially classes 1 and 2 are underrepresented, whereas class 0 is overrepresented. These could lead to biased model learning. Therefore, the classes were all undersampled to the minority class. Finally, for the MDB, the training data included 818489 windows for each class, the validation data included 177582, and the test data included 179547 windows for each class. In SDB II, the respective splits included 4364, 923, and 1083 windows per class for training, validation, and testing.

\subsection{Model Architecture}

Fawaz et al. provided a comprehensive review of state-of-the-art time series classification models and highlighted the superiority of the ResNet architecture proposed by Wang et al. \cite{fawaz_deep_2019, wang_time_2016}. For benchmarking hypoglycemia classification, we applied the same ResNet model without modifying its architecture, except for replacing 2DCNN layers with 1DCNN layers. The model of Wang et al. comprises three ResNet blocks, each containing three sequential 1DCNN layers with kernel sizes of 8, 5, and 3, respectively. Each convolutional layer is followed by a batch normalization and a ReLU activation function. No dropout or dense layers are used \cite{fawaz_deep_2019, wang_time_2016}. After comparing the baseline model with a variant incorporating the "He Normal" kernel initializer, the latter was used for training due to better performance and shorter computation time. A batch size of 256 was selected to accommodate the large dataset, which included approximately 4 million windows. The training was conducted for a maximum of 50 epochs, with early stopping applied if validation accuracy did not improve for 3 consecutive epochs. The initial learning rate of 0.0001 was reduced if no improvement was observed over 5 epochs. The batch size of SDB II was set to 64.

To enhance training efficiency, input and output values were converted to TensorFlow-compatible tensors, and the training data was shuffled. Labels were kept as integer values rather than one-hot encoded to reduce computational complexity and training time. Accordingly, sparse categorical cross-entropy and sparse categorical accuracy were used as the loss function and evaluation metric, respectively. A population-based model was trained with a hold-out validation approach.

All experiments were performed on a MacBook Pro 2023 with 32 GB RAM and an Apple M2 chip using its integrated GPU, running TensorFlow version 2.13.0 and Python 3.11.4.

\section{Results}
\label{sec:results}
This section presents the result of the data analysis process with the quality-enhanced DiaData, consisting of a correlation analysis between sensor values and a time series classification model to predict the onset of hypoglycemia.

\subsection{Correlation Analysis}

The correlation between glucose and heart rate in SDB II was estimated using Spearman's rank correlation, yielding an overall low coefficient score of 0.065. Furthermore, correlation values specific to hypoglycemic classes are presented in Table \ref{tab:corrhr}. Notably, a moderate positive correlation between glucose and heart rate is observed 15 to 60 minutes before hypoglycemia, whereas the correlation weakens at the exact time of the event. Class 5, representing data points more than two hours before hypoglycemia or unrelated to hypoglycemic events, indicates no significant correlation. These findings suggest that heart rate data provide relevant temporal information in lower glucose ranges, potentially enhancing model performance and improving class differentiation.
\begin{table}[h!]
\centering
\caption{Spearman's Rank Correlation of Sensor Data}
\label{tab:corrhr}
\begin{tabular}{|c|c|c|c|c|c|c|}
\hline
\textbf{Class} & \textbf{0} & \textbf{1} & \textbf{2} & \textbf{3} & \textbf{4} & \textbf{5}  \\
\hline 
\textbf{Correlation} &0.117 & 0.195 &  0.300 & 0.267 & 0.205 & 0.084  \\ 
\hline
\end{tabular}
\end{table}

\subsection{Model Results}

This subsection reports the performance of ResNet models in classifying hypoglycemia up to 2 hours before onset. Models were independently trained and evaluated on the quality-refined Maindatabase (MDB) and Subdatabase II (SDB II), as well as corresponding raw datasets. In the raw dataset, outliers and missing values were not treated, whereas parameters were normalized, and class imbalance was mitigated to ensure fair comparison. Average (Avg) and class-wise performance metrics for all models are summarized in Table \ref{tab:Comp5classles}. 
\begin{table}[!ht]
\centering
\caption[Performance of ResNet Model]{Performance of ResNet Model\label{tab:Comp5classles}}
\begin{tabular}{c|c|c|c|c|c|c|c|}
\cline{2-8}
    & \textbf{Metric} & \textbf{Avg} &
    \multicolumn{5}{ l |} {\textbf{Class}}\\ 
    \cline{4-8} & & & 0& 1  & 2 & 3 & 4  \\ \hline

\multicolumn{1}{|l|}{\multirow{4}{*}{}} & Precision & 0.74 & 1.00 & 0.83 & 0.62& 0.54 &  0.70 \\  

\cline{2-8} 
\multicolumn{1}{|l|}{\textbf{MDB}} & Recall  & 0.74 & 0.99 & 0.85& 0.66 & 0.51 & 0.67 \\ 

\cline{2-8}
\multicolumn{1}{|l|}{}  & F1-M & 0.74& 1.00& 0.84&0.64&0.52 & 0.69 \\   
 \hline

\multicolumn{1}{|l|}{\multirow{4}{*}{}} & Precision & 0.71 & 1.00 & 0.78 & 0.57& 0.52 &  0.69 \\  

\cline{2-8} 
\multicolumn{1}{|l|}{\textbf{Raw MDB}} & Recall  & 0.72 & 1.00 & 0.84& 0.62 & 0.46 & 0.67  \\ 

\cline{2-8}
\multicolumn{1}{|l|}{}  & F1-M & 0.71& 1.00& 0.81&0.59&0.49 & 0.68 \\   
 \hline

\multicolumn{1}{|l|}{\multirow{4}{*}{}} &  Precision & 0.66& 0.96& 0.71&0.51&0.48 & 0.63 \\    

\cline{2-8}  
\multicolumn{1}{|l|}{\textbf{SDBII}} & Recall  &0.66 & 0.92& 0.75&0.55&0.45 & 0.60  \\

\cline{2-8} 
\multicolumn{1}{|l|}{}  & F1-M  & 0.66&0.94 &0.73 &0.53& 0.46& 0.61  \\
\hline

\multicolumn{1}{|l|}{\multirow{4}{*}{}} &  Precision & 0.59& 0.93& 0.60&0.39&0.40 & 0.64 \\    

\cline{2-8}  
\multicolumn{1}{|l|}{\textbf{Raw SDBII}} & Recall  &0.59 & 0.92& 0.58&0.43&0.37 & 0.65  \\

\cline{2-8} 
\multicolumn{1}{|l|}{}  & F1-M  & 0.59&0.93&0.59 &0.41 &0.38& 0.65 \\
\hline
\end{tabular}
\end{table}

The results underscore the impact of data quality, yielding a 2-3\% increase in the performance of the MDB compared to the raw dataset, particularly for classes 1, 2, 3, and 4. SDB II shows an improvement of 7\%.

Moreover, the quality-refined MDB, representing a larger and more heterogeneous population limited to CGM data only, achieves approximately 8\% higher performance across all metrics compared to the quality-refined SDB II. This highlights the substantial advantage of data quantity, revealing that a larger univariate dataset outperforms a smaller multivariate dataset, even when the latter demonstrates moderate levels of correlation between sensor values. Hypoglycemic events (class 0) were classified with high accuracy in both datasets. However, the larger dataset yielded significantly better F1-M scores for the remaining classes. Although class 3 performance improves by 6\%, the metrics remain below 55\%, indicating a high rate of misclassifications.  
The models did not overfit and showed no discrepancies between training, validation, and test performance, indicating good generalization to unseen data. However, both quality-improved datasets show similar classification trends, and cannot classify all classes with high confidence, as also reported in prior studies \cite{cinarMaster, xie_transfer_2024}.

\begin{figure}[hb]
\centering
\begin{subfigure}{0.24\textwidth}
    \includegraphics[width=\textwidth]{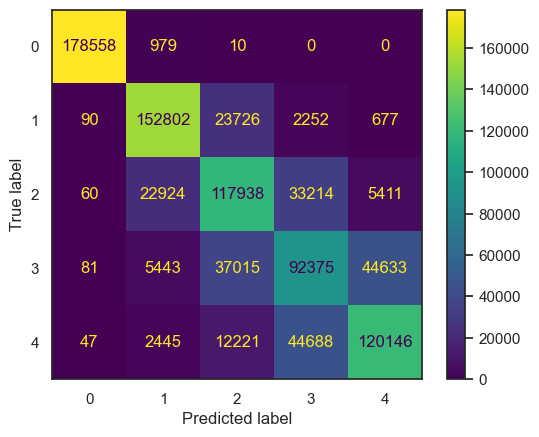}
    \caption{Maindatabase}
    \label{fig:cmMain}
\end{subfigure}
\begin{subfigure}{0.23\textwidth}
    \includegraphics[width=\textwidth]{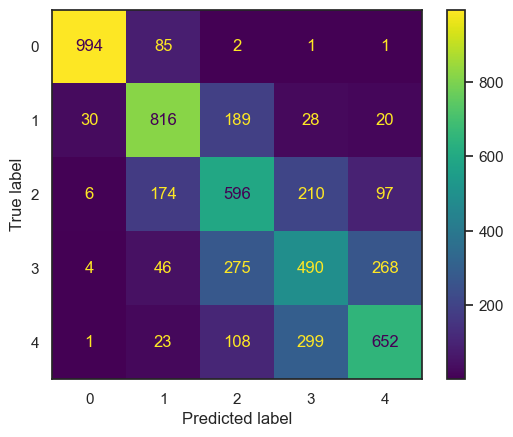}
    \caption{Subdatabase II}
    \label{fig:cmSub}
\end{subfigure}
\caption{Confusion Matrices}
\label{fig:CM}
\end{figure}

Corresponding confusion matrices presented in Fig. \ref{fig:CM} show that misclassifications predominantly occurred between adjacent classes. For instance, class 1 was often misclassified as class 2, implying that alerts would occur earlier. Thus, the event is still detected but with a shifted time. Class 2 was typically misclassified as class 1 or 3, class 4 as class 3, and class 4 showed the weakest performance, highlighting a need for further improvement. These findings align with prior research, which similarly reported declining classification stability with increasing PHs. 
Conclusively, the importance of data quality and quantity is highlighted, providing better generalization and increased performance. The correlation analysis suggests that additional context and sensor values can enhance classification performance, particularly in distinguishing between increased PHs \cite{Felizardo2021review}. This is also supported by the observation that even the smaller dataset exhibited similar patterns and results.

\section{Discussion}
\label{sec:discussion}
This study preprocessed DiaData, a large CGM database of subjects with T1D, introduced to address the lack of publicly available datasets for T1D research. To support reliable prediction models, we enhanced data quality through anomaly detection and missing value imputation. First, outliers that distorted sensor data with non-realistic values were masked and replaced with missing values. Preliminary statistical analysis for glucose and heart rate data revealed several values outside physiologically realistic ranges. Retaining these outliers may lead to overestimation or degraded performance since the model can become biased by noise and misleading sensor errors. Therefore, we applied the IQR method to remove values outside the valid ranges. Heart rate data revealed outliers in lower ranges, whereas zero values were included in the first quartile of data in CGM values, marking a considerable number of outliers in hypoglycemic ranges. In both sensor values, a high volume of outliers in higher ranges was recorded. Thus, anomaly treatment showed improvement in the variation of both parameters and was particularly effective for the D1NAMO and T1GDUJA datasets. 

Subsequently, missing values were imputed to address numerous short-term gaps that could otherwise lead to the loss of relevant time windows. These gaps likely result from random sensor errors or undersampling in certain datasets. Short-term gaps can be addressed effectively using appropriate imputation methods as reported in section \ref{sec:stoa}. Different imputation methods were applied based on gap length, as linear interpolation may not accurately capture glucose dynamics over longer gaps \cite{GuptaRev2025titlex}. Visual analysis indicated that Stineman interpolation, which uses smooth polynomial slopes, provided a more realistic approximation of CGM values. It was noticed that some short- and middle-term gaps remained, possibly because not enough values between the gaps were present. These may be addressed with different imputation approaches in future work. While Stineman interpolation realistically reflects glucose patterns for mid-term gaps up to 2 hours, more advanced ML-based imputation methods should be explored for longer gaps of 4h, which can significantly affect the performance of ML models \cite{Rehman2024}. In addition, the same methods were applied to heart rate data. While linear interpolation remains suitable for short gaps, more advanced techniques should be considered for gaps longer than 30 minutes, as Stineman interpolation may not adequately capture heart rate dynamics \cite{GuptaRev2025titlex, Gupta2025-hk}.

Furthermore, a benchmark for hypoglycemia classification was conducted using the quality-refined and raw Maindatabase and the Subdatabase II. Results demonstrated a 2-3\% performance gain from quality improvement, supporting prior findings, and highlighting the impact of data quantity with an increase of 7\%, despite using univariate data. 
Although the quality-refined datasets differ in size, they yield similar classification patterns. Moreover, the subset including heart rate data shows potential benefits due to its moderate correlation with glucose levels 15–60 minutes before hypoglycemia. 
DiaData outperformed our previous model trained with the OhioT1DM dataset \cite{cinarMaster}, and produced results comparable to those in Table~\ref{tab:CompStoA}, with Dave et al. achieving better outcomes using binary classification. Unlike other studies, our metrics were well-balanced. However, our model could not classify well between 30 min and 1h before hypoglycemia. 

Future work should explore more stable architectures, including ensemble methods and the incorporation of demographic features. ML may perform better on smaller datasets, while complex DL models may be more suitable for large datasets, aiming at population-based models. In addition, some subjects may have introduced bias. Specifically, some individual datasets within the integrated DiaData have larger data volumes compared to others \cite{DiaData_Preprint}. To mitigate potential bias, these subjects could be excluded or their data volume could be reduced. Alternatively, they may be used for model individualization.

\section{Conclusion}
\label{sec:conclusion}
This study improved the quality of DiaData, a large database of 2510 patients with T1D, containing CGM values measured every 5 min. Outliers were identified with the IQR method and treated as missing values, which led to less bias toward misleading sensor errors. Then, missing data were categorized by gap length of less than 30 min, between 30 and 120 min, and larger than 120 min. Small gaps were imputed with linear interpolation, while gaps of the second category were imputed with Stineman interpolation, providing a more natural glucose curve.
Furthermore, this study presents a benchmark on hypoglycemia classification. The model classified the time range before hypoglycemia onset. For preprocessing, classes were assigned, sensor data were normalized, time series were created, and class imbalance was mitigated through undersampling. Data analysis on the preprocessed DiaData revealed a moderate correlation between glucose and heart rate 15 to 60 minutes before hypoglycemia. Finally, a state-of-the-art ResNet model trained on DiaData achieved high confidence predictions up to 15 min before hypoglycemia, and moderate confidence up to 30 min, whereas up to 2 hours were often misclassified. The quality refinement led to an increased performance by 2-3\%. While heart rate is a valuable parameter, training on a large dataset with only CGM values improved performance by 7\% with a total F1-Measure of 74\%. Future work should explore advanced imputation methods for both sensor types separately, better model architectures, and contextual data to improve the classification performance of increased PHs.

\subsection*{Data and Code Availability}

The datasets used in this study were obtained from multiple third-party sources. A subset of the data can be downloaded from the following source: \href{https://zenodo.org/records/16875703}{https://zenodo.org/records/16875703} \cite{DiaDataZenodo}. All data integration, cleaning, and preprocessing scripts, as well as high-quality versions of included illustrations, are publicly available at the following GitHub repository: \href{https://github.com/Beyza-Cinar/Preprocessing-DiaData}{https://github.com/Beyza-Cinar/Preprocessing-DiaData}. 

The sources of subsets of the data are the Barbara Davis Center, Jaeb Center for Health Research, Joslin Diabetes Center, T1D Exchange, University of Colorado, and the University of Virginia. Retrieved from: https://public.jaeb.org/dataset. The analyses, content, and conclusions presented herein are solely the responsibility of the authors and have not been reviewed or approved by the aforementioned institutions.

\bibliographystyle{ieeetr}
\bibliography{Literature}

\end{document}